# NMR signals within the generalized Langevin model for fractional Brownian motion

Vladimír Lisý [a, b], Jana Tóthová [a]

[a] *Department of Physics, Technical University of Košice,
Park Komenského 2, 042 00 Košice, Slovakia*
[b] *Laboratory of Radiation Biology, Joint Institute of Nuclear Research,
141 980 Dubna, Moscow Region, Russia*

**Abstract.** The methods of Nuclear Magnetic Resonance belong to the best developed and often used tools for studying random motion of particles in different systems, including soft biological tissues. In the long-time limit the current mathematical description of the experiments allows proper interpretation of measurements of normal and anomalous diffusion. The shorter-time dynamics is however correctly considered only in a few works that do not go beyond the standard memoryless Langevin description of the Brownian motion (BM). In the present work, the attenuation function $S(t)$ for an ensemble of spin-bearing particles in a magnetic-field gradient, expressed in a form applicable for any kind of stationary stochastic dynamics of spins with or without a memory, is calculated in the frame of the model of fractional BM. The solution of the model for particles trapped in a harmonic potential is obtained in an exceedingly simple way and used for the calculation of $S(t)$. In the limit of free particles coupled to a fractal heat bath, the results compare favorably with experiments acquired in human neuronal tissues. The effect of the trap is demonstrated by introducing a simple model for the generalized diffusion coefficient of the particle.

## 1. Introduction

Nuclear Magnetic Resonance (NMR) has proven to be a very effective non-invasive method of studying molecular self-diffusion and diffusion in various materials [1–6]. In some complex media particles often exhibit interesting behaviors, such as an "anomalous" dynamics that significantly differs from that predicted by the standard Einstein [7] and Langevin [8] theory of the Brownian motion (BM) [9–12]. It is thus natural to apply the NMR methods to study such anomalous BM. When the particles are in the diffusion regime, i.e., at long times (the times much larger than the characteristic frictional time of the particles [9]), such studies have really been presented in a number of papers, e.g. [13–27]. As discussed in [28–31], the mathematical description of the NMR experiments in the literature is not valid for the all-time dynamics of the spin-bearing particles. The only exception is the simplest memoryless Langevin model for which correct interpretations of the NMR experiments have been given, first in [32]. Possible memory effects in the shorter-time particle dynamics are ignored or incorrectly taken into account. The aim of our recent papers [28–31] was to overcome this limitation and to calculate the all-time attenuation functions of the NMR signals from spin-bearing particles resulting from the BM with memory. We have considered the Brownian particles in Maxwell fluids [28, 29] and the BM with hydrodynamic memory [30]. In the present work we focus on a popular fractional BM model for the description of transport dynamics in complex fluids and different physical, chemical and biological systems [33–38], but also for open quantum systems, econophysics [39] and medicine [40]. In the condensed matter physics this model emerges naturally, e.g., in viscoelastic media, and interpolates between the standard Langevin equation (LE) with the white noise force and the purely viscous Stokes friction force, and a model with constant memory [38]. Our aim in this paper is to describe the all-time dynamics of the particles and how it can be probed by the NMR methods. We obtain the solution of the generalized Langevin equation (GLE) for this model in an exceedingly simple way. The Brownian particles can be free or trapped in a harmonic potential. The results for the mean square displacement (MSD) of the particles are then applied to calculate the attenuation function $S(t)$ for an ensemble of spins in a magnetic-field gradient. $S(t)$ is used in the form previously obtained by the method of accumulation of phases in the frame rotating with the resonance frequency [29] and applicable for any time and kind of the stochastic motion of spins. Although our work was mainly aimed to contribute to adequate interpretations of the NMR experiments on the BM of large particles in complex fluids, in the limit of free particles coupled to a fractal heat bath our results correct the description and give a favorable comparison with experiments acquired in human neuronal tissues [23].

## 2. Harmonically bounded particle described by the fractional Langevin equation

If a particle is moving in a trap modeled by a Hookean potential with elastic constant $k$, the GLE reads [41]

$$M\dot{\upsilon}(t)+\int_0^t \Gamma(t-t')\upsilon(t')dt'+k\int_0^t \upsilon(t')dt' = f(t), \tag{1}$$

where $M$ is the mass of the particle, $x(t)$ is its position and $\upsilon(t) = \dot{x}(t)$ its velocity. We will consider the memory kernel $\Gamma(t) = \gamma_\varepsilon \varepsilon t^{\varepsilon-1}$, $t > 0$, $1 > \varepsilon > 0$ [37]. Within the linear response theory the second fluctuation-dissipation theorem [41] relates $\Gamma(t)$ to the random force $f(t)$, $\langle f(t)f(0)\rangle = k_B T\Gamma(t)$, $t > 0$. When $\varepsilon \to 0$, $\Gamma(t) \to 2\gamma\delta(t)$ and Eq. (1) becomes the standard memoryless LE with the white noise force and the purely viscous Stokes friction force $-\gamma\upsilon(t)$. To our opinion, the most simple method of solving the GLE equations, which goes back to the old work by Vladimirsky [42, 43], is as follows [44–48]. If we are interested in finding the mean square displacement (MSD) of the particle, $X(t) = \langle \Delta x^2(t) \rangle = \langle [x(t) - x(0)]^2 \rangle$, we have merely to replace $\upsilon(t)$ in (1) by $V(t) = dX(t)/dt$ and to substitute the stochastic force driving the particle with $2k_B T$, where $k_B$ is the Boltzmann constant and $T$ the temperature. The equation of motion must be solved with the initial conditions $X(0)=V(0)=0$. Obviously, also the condition $\dot{V}(0)=2k_BT/M$ must hold. For more details see the above cited articles [42–48]. The new equation,

$$M\ddot{X}(t)+\int_0^t \Gamma(t-t')\dot{X}(t')dt'+kX(t)=2k_BT, \qquad (2)$$

is immediately solved by using the Laplace transformation,

$$\tilde{X}(s)=\int_0^\infty X(t)e^{-st}dt=\frac{2k_BT}{Ms}\left[s^2+\frac{\gamma_\varepsilon}{M}\Gamma(\varepsilon+1)s^{1-\varepsilon}+\frac{k}{M}\right]^{-1}, \qquad (3)$$

where $\Gamma(z)$ is the gamma function [49]. Thus, in this way the exact solution of Eq. (1) [37], can be easily got in a few steps. The limiting cases of short and long times correspond to large and small $s$, respectively, so that if $\varepsilon<1$ than at $s\to 0$ the main approximation of (3) will be $\tilde{X}(s)\simeq 2k_BT/ks$ and $X(t)\simeq 2k_BT/k$, if $t\to\infty$ [49]. At $s\to\infty$, $\tilde{X}(s)\simeq 2k_BT/Ms^3$ and $X(t)\simeq k_BTt^2/M$. Corrections to these expressions can be obtained from the exact representation of Eq. (3) in the time domain through the Mittag-Leffler functions $E_{\alpha,\beta}(y)$ [37],

$$X(t)=\frac{2k_BT}{M}\sum_{k=0}^\infty \frac{(-1)^k}{k!}(\omega t)^{2k} t^2 E_{1+\varepsilon,3+k-\varepsilon k}^{(k)}\left(-\varepsilon\gamma_\varepsilon\Gamma(\varepsilon)t^{1+\varepsilon}/M\right), \qquad (4)$$

where $E_{\alpha,\beta}(y)=\sum_{j=0}^\infty y^j \Gamma^{-1}(\alpha j+\beta)$, $\alpha>0$, $\beta>0$, and $E_{\alpha,\beta}^{(k)}(y)=d^k E_{\alpha,\beta}(y)/dy^k$. It should be noted here that the short-time expansion of this exact solution for $X(t)$, as given in [37] (Eq. (41)), does not correspond to the solution of the standard Langevin equation (Eq. (1) in the limit $\varepsilon\to 0$) [50]

$$X(t)=\frac{2k_BT}{M}\frac{1}{\lambda_2-\lambda_1}\left(\frac{1}{\lambda_1}-\frac{1}{\lambda_2}+\frac{\exp(\lambda_2 t)}{\lambda_2}-\frac{\exp(\lambda_1 t)}{\lambda_1}\right), \qquad (5)$$

with $\lambda_{1,2}$ being the roots of the equation $\lambda^2+\gamma\lambda/M+k/M=0$, $2\lambda_{1,2}=-\omega_M[1\mp(1-4\omega^2\omega_M^{-2})^{1/2}]$, where $\omega^2=k/M$, $\omega_M=\gamma/M$ and $\lambda_1\lambda_2=\omega^2$. In [37], Eq. (41), the term $\sim t^4$ in $\{.\}$ should be $(1/12)(\omega_M^2-\omega^2)t^4$, i.e., the term containing $\omega_M$ is missing. Another way to show that the expansions in [37] should be completed is to consider the limit of the free Brownian particle (when $\omega = 0$). The velocity autocorrelation function in this case must be $\langle v(0)v(t)\rangle=(k_BT/M)\exp(-\omega_M t)$, which does not agree with Eq. (42) [37]. The expansion that gives the correct simplification to the solution of the standard LE is for $C_v(t)=\langle v(0)v(t)\rangle/\langle v^2\rangle$

$$C_\upsilon(t) = 1 - \frac{\gamma_\varepsilon}{M(1+\varepsilon)} t^{1+\varepsilon} + \left(\frac{\gamma_\varepsilon}{M}\right)^2 \frac{\Gamma^2(1+\varepsilon)}{\Gamma(3+2\varepsilon)} t^{2+2\varepsilon} - \frac{1}{2}(\omega t)^2 + \ldots, \qquad (6)$$

and the MSD should be

$$\frac{M}{k_B T} X(t) = t^2 - 2\frac{\gamma_\varepsilon}{M}\frac{\Gamma(1+\varepsilon)}{\Gamma(4+\varepsilon)} t^{3+\varepsilon} + 2\left(\frac{\gamma_\varepsilon}{M}\right)^2 \frac{\Gamma^2(1+\varepsilon)}{\Gamma(5+2\varepsilon)} t^{4+2\varepsilon} - \frac{\omega^2 t^4}{12} + \ldots, \qquad (7)$$

In the case of a free particle, the solution of the GLE takes the form [37]

$$X(t) = \frac{2k_B T}{M} t^2 E_{\varepsilon+1,3}\left(-\frac{\gamma_\varepsilon}{M}\Gamma(\varepsilon+1) t^{\varepsilon+1}\right). \qquad (8)$$

This equation was used by Cooke et al. [23] in the calculation of the attenuation function $S(t)$ of the NMR signals in diffusion-weighted experiments on human neuronal tissues. Below it will be shown that this calculation must be corrected in several points. As a consequence, the final result for the spin-echo signal proposed in [23] must be changed. $S(t)$ will be also obtained for bounded spin-bearing particles, whose MSD is described by Eq. (3) and its short- and long-time approximations in the time domain. The basic formulas used to calculate $S(t)$ are presented in the following section.

### 3. Attenuation of the NMR signals due to Brownian motion

Below we will consider experiments, in which the NMR signals are taken from a liquid or gaseous system with spin-bearing Brownian particles placed in a strong magnetic field $B_0$ along the axis $x$. The magnetization of spins is modulated by a field gradient $g(t)$. General formulas for the attenuation function $S(t)$ of the observed NMR signal that are applicable to any kind of stochastic motion of spins, including their dynamics with memory, have been obtained in [28, 29, 31]. In the evaluation of $S(t)$ it was assumed that the studied random processes are stationary in the sense that the autocorrelation function of a fluctuating dynamical variable $x$ at times $t$ and $t'$ depends only on the time difference $t - t'$: $\langle x(t)x(t')\rangle = \langle x(t-t')x(0)\rangle$ [6]. One more assumption is that the distribution of the random variables is Gaussian, or that the accumulated phase $\phi$ in the frame rotating with the resonance frequency $\gamma_n B_0$, where $\gamma_n$ is the nuclear gyromagnetic ratio, is small. In these cases the nuclear induction signal observed in the presence of a steady magnetic-field gradient after the 90° rf pulse applied at time $t = 0$, $S(t) = \langle \exp[i\phi(t)]\rangle$, is [1–3, 5, 6, 51]

$$S(t) = \exp\left[-\frac{1}{2}\langle \phi^2(t)\rangle\right]. \qquad (9)$$

The phase accumulation in the rotating frame in Eq. (9) must be calculated through the change of the phase during the time $t$, which gives

$$\phi(t) = -\gamma_n \int_0^t \left[ B(x(t')) - B(x(0)) \right] dt' = -\gamma_n \int_0^t g\left[ x(t') - x(0) \right] dt', \tag{10}$$

where the Larmor condition $\omega = -\gamma_n B$ was used, and $B = B_0 + g(t)x(t)$. Assuming a constant gradient during the interval $t$, it follows from (10) and (9)

$$S(t) = \exp\left[ -\frac{1}{2} \gamma_n^2 g^2 \int_0^t t' X(t') dt' \right]. \tag{11}$$

To our knowledge, this simple formula was for the first time proposed in [28].

In the pulsed-gradient Hahn echo experiment [52], between two gradient pulses, each of duration $\delta$, a 180° rf pulse is applied. Let the first gradient pulse begins at time $t = t_0$ after the 90° rf pulse and the second one at time $t = \Delta$ [6]. The result for $t < \tau$ (up to the second rf pulse) is the same as for a steady gradient (Eq. (11) with $t = \delta$). Due to stationarity, after the second rf and gradient pulses the result of the calculation of the echo signal observed at $t = 2\tau$ does not depend on $t_0$ and can be evaluated from [29]

$$S(\delta, \Delta) = \exp\left\{ -\frac{1}{2} \gamma_n^2 g^2 \left[ \int_0^\delta dt' \int_0^\delta dt'' X(t'' - t' + \Delta) - 2\int_0^\delta dt' (\delta - t') X(t') \right] \right\}. \tag{12}$$

This formula simplifies to the well-known relation $S(\delta, \Delta) = \exp[-\gamma_n^2 g^2 D \delta^2 (\Delta - \delta/3)]$ [3] when the MSD is $X(t) = 2Dt$. If $\delta = \Delta = \tau$ is substituted in (12), we obtain the damping of the signal at the echo time $2\tau$, in the case of the steady-gradient echo [53].

Equations (11) and (12) have been applied in [29] to calculate the attenuation function in the case of the BM in viscoelastic fluids, when it is described by the GLE with the friction force modeled by the convolution of an exponentially decaying memory kernel [54, 55]. In particular, it has been shown that the correction to the above given result for $S(\delta, \Delta)$ based on the Einstein theory of diffusion can be significant. This depends on the relation between the durations of the gradient pulses and the interval $\Delta$. If at long times the particles exhibit anomalous diffusion, $X(t) = Ct^\alpha$, where $C$ is a temperature-dependent parameter, $\alpha = 1$ corresponds to normal diffusion with $C = 2D$ with the diffusion coefficient $D = k_B T / \gamma$, $\alpha < 1$ to sub-diffusion, and $\alpha > 1$ to super-diffusion [56, 45], Eqs. (11) and (12) allows to easily obtain the attenuation, which is very different from the result for the fractional diffusion, found in [23]. Below we extend this consideration also for short times and consider the fractional BM of free and harmonically bounded particles, based on the solutions for the MSDs given in the preceding section.

## 4. Free particle in the fractional Brownian motion

In Ref. [23], in the case of the steady gradient, the solution (8) was used to calculate $\langle \phi^2(t) \rangle$ for $S(t)$ in Eq. (9) from the expression

$$\langle \phi^2(t) \rangle = 2\gamma_n^2 g^2 \int_0^t \int_0^t t_1 t_2 \langle \dot{x}(t_1) \dot{x}(t_2) \rangle dt_1 dt_2, \tag{13}$$

which is correct (although more complicated than that used in our Eq. (11)) but its derivation is wrong. Cooke et al. [23] apply the equation $\phi^2(t) = 2\int_0^t \phi(t_1) \dot{\phi}(t_1) dt_1$. They get $\phi(t) = -\gamma_n \int_0^t \int_0^{t_1} dt_1 dt_2 \dot{x}(t_1) g(t_2)$, which is not correct. In its derivation by integration by parts from the equation $\phi(t) = \gamma_n \int_0^t x(t_1) g(t_1) dt_1$ they impose the "rephasing condition" $\int_0^t g(t') dt' = 0$, which evidently does not hold for the steady gradient. Nevertheless, when using the correct expression $\phi(t) = \gamma_n g t x(t) - \gamma_n g \int_0^t \dot{x}(t_1) t_1 dt_1$ (for a constant gradient), the correct result (13) is obtained. This is because the term $\gamma_n g t x(t)$ in $\phi(t)$ does not contribute to $\langle \phi^2(t) \rangle$. However, the calculation of $\langle \phi^2(t) \rangle$ for the spin echo in [23], Eq. (43), the details of which are given in Appendix B, is wrong. It is seen from the long-time dependence of $\langle \phi^2(t) \rangle$, which must be the same as that for anomalous diffusion with the MSD $\langle X(t) \rangle = Ct^\alpha$. The correct $\langle \phi^2(t) \rangle$ for this case [15] (see below Eq. (16) with $\alpha = 1 - \varepsilon$) significantly differs from the result in [23]. In the calculation of $\langle \Delta \phi^2(t) \rangle$ for the spin echo they consider four intervals, viz., $(0, \delta)$, $(\delta, \Delta)$, $(\Delta, \Delta + \delta)$, $(\Delta + \delta, 2\tau)$, and evaluate the contributions to $\langle \phi^2(t) \rangle$ separately for all the intervals. In spite of the fact that in the second and fourth interval the gradient is absent, the contributions are found to be $\sim g$. The summation then gives the incorrect result, Eq. (43) [23]. Here, instead of (13), we use a more simple expression (11). With this equation and the solution (8), the attenuation function is straightforwardly calculated. By using the expansion of $E_{\alpha,\beta}(y)$ shown after Eq. (4), the $t \to 0$ behavior of $S(t)$ is given mainly by the formula $S(t) \approx \exp[-\gamma_n^2 g^2 k_B T t^4 / 8M]$, which is the same as in the standard Langevin and other models. The correction to $\langle \phi^2(t) \rangle$ in the considered model is given by the equation

$$\langle \phi^2(t) \rangle = \frac{2\gamma_n^2 g^2 k_B T}{M} \left( \frac{t^4}{8} - \frac{\gamma_\varepsilon \Gamma(\varepsilon+1)}{M(\varepsilon+5)\Gamma(\varepsilon+4)} t^{\varepsilon+5} + ... \right). \tag{14}$$

At $\varepsilon = 0$ it agrees with the standard Langevin model, for which the MSD is $X(t) = 2D\{t - (M/\gamma)[1 - \exp(-\gamma t/M)]\}$.

To obtain the long-time limit at $\gamma_\varepsilon \Gamma(\varepsilon+1) t^{\varepsilon+1}/M \gg 1$, one can use [35–37] $E_{\alpha,\beta}(-z) \sim [z\Gamma(\beta-\alpha)]^{-1}$, $z > 0$, and the properties of the gamma function [49] $\Gamma(z+1) = z\Gamma(z)$ and $-z\Gamma(z)\Gamma(-z) = \pi \csc \pi z$. The MSD is then approximated by the formula $X(t) \approx C_\varepsilon t^{1-\varepsilon}$ with $C_\varepsilon/2k_BT = [\gamma_\varepsilon \Gamma(2-\varepsilon)\Gamma(1+\varepsilon)]^{-1} = [\gamma_\varepsilon(1-\varepsilon)\pi\varepsilon]^{-1} \sin \pi\varepsilon$. Thus, for the steady gradient at long times

$$\langle \phi^2(t) \rangle \approx \frac{C_\varepsilon \gamma_n^2 g^2}{3-\varepsilon} t^{3-\varepsilon}. \tag{15}$$

In the case of the pulsed-gradient echo, the decrease of the NMR signal at any time after the second gradient pulse will be from Eq. (12)

$$\langle \phi^2(\delta,\Delta) \rangle \approx \frac{C_\varepsilon \gamma_n^2 g^2}{(2-\varepsilon)(3-\varepsilon)} \left[ (\Delta+\delta)^{3-\varepsilon} + (\Delta-\delta)^{3-\varepsilon} - 2\Delta^{3-\varepsilon} - 2\delta^{3-\varepsilon} \right]. \tag{16}$$

This equation corresponds to the result found in [15] but significantly differs from that by Cooke et al., Eq. (43) [23], where $\langle \phi^2(\delta,\Delta) \rangle \propto \delta^2 [(\Delta-\delta)^{1-\varepsilon} + 2\delta^{1-\varepsilon}/(3-\varepsilon)]$.

## 5. Particle in a trap

For the steady-gradient experiment we obtain from Eqs. (9) and (11) with the MSD (7)

$$\langle \phi^2(t) \rangle = \frac{\gamma_n^2 g^2 k_B T}{M} \left( \frac{t^4}{4} - \frac{2\gamma_\varepsilon \Gamma(1+\varepsilon)}{M(5+\varepsilon)\Gamma(\varepsilon+4)} t^{5+\varepsilon} + \left(\frac{\gamma_\varepsilon}{M}\right)^2 \frac{\Gamma^2(1+\varepsilon)}{(3+\varepsilon)\Gamma(5+2\varepsilon)} t^{6+2\varepsilon} - \frac{\omega^2}{72} t^6 + \ldots \right). \tag{17}$$

The expansion is done up to the term $\sim t^6$ corresponding to the fact that the particle begins to "feel" the trap. At $\varepsilon = 0$ this formula agrees with that from the solution (5) of the standard LE for a trapped particle, and at $\omega = 0$ we have the result for a free particle.

The echo attenuation obtained from Eq. (12) is at short times

$$\langle \phi^2(\delta,\Delta) \rangle = \frac{\gamma_n^2 g^2 k_B T}{M} \left( \frac{1}{12} \varphi_4 - 2 \frac{\gamma_\varepsilon}{M} \frac{\Gamma(1+\varepsilon)}{\Gamma(6+\varepsilon)} \varphi_{5+\varepsilon} + 2\left(\frac{\gamma_\varepsilon}{M}\right)^2 \frac{\Gamma^2(1+\varepsilon)}{\Gamma(7+2\varepsilon)} \varphi_{6+2\varepsilon} - \frac{\omega^2}{360} \varphi_6 + \ldots \right), \tag{18}$$

with $\varphi_\alpha = (\Delta+\delta)^\alpha + (\Delta-\delta)^\alpha - 2\Delta^\alpha - 2\delta^\alpha$. At $\delta \ll \Delta$ it holds $\varphi_\alpha \approx \alpha(\alpha-1)\Delta^{\alpha-2}\delta^2$ and one obtains $\langle \phi^2(\delta,\Delta) \rangle = (\gamma_n g \delta)^2 X(\Delta)$, with $X(t)$ from (7). It is easy to return to the solution based on the standard Langevin theory for both the trapped ($\varepsilon = 0$) and free particles ($\varepsilon = 0$, $\omega = 0$).

The long-time approximation needs more attention. In the work [37], Eq. (48), the MSD obtained for $t^{1-\varepsilon} \gg \gamma_\varepsilon \Gamma(1+\varepsilon)/M\omega^2$,

$$X(t) \approx \frac{2k_B T}{M\omega^2}\left[1 - \frac{\varepsilon \gamma_\varepsilon}{M\omega^2} t^{-1+\varepsilon}\right], \quad (19)$$

cannot be used to get the $\omega \to 0$ limit and the standard Langevin limit with a trap ($\varepsilon = 0$) gives only the main approximation $X(t) \approx 2k_B T / M\omega^2$. To obtain the long-time behavior valid also for the vanishing trapping force one must use the solution (4) in terms of the Mittag-Leffler functions, first by introducing at $\gamma_\varepsilon \Gamma(1+\varepsilon) t^{1+\varepsilon} / M \gg 1$ the asymptotic behavior of these functions. This allows expressing the MSD as

$$X(t) \approx \frac{2k_B T}{M\omega^2}\left[1 - E_{1-\varepsilon}\left(-\frac{\omega^2 M t^{1-\varepsilon}}{\gamma_\varepsilon \Gamma(1+\varepsilon)}\right)\right], \quad (20)$$

where $E_\alpha(z) = E_{\alpha,1}(z)$ is a one-parameter Mittag-Leffler function. At $\varepsilon = 0$, by using $E_1(z) = \exp(z)$, it follows from this equation the MSD within the standard Langevin model, $X(t) \approx 2k_B T[1 - \exp(-\omega^2 M t / \gamma)] / M\omega^2$, $\gamma t / M \gg 1$ [50]. At $\omega \to 0$ this gives the correct behavior at long times, $X(t) \approx 2k_B T t / \gamma$. For arbitrary $\varepsilon$ and $M\omega^2 t^{1-\varepsilon} \ll \gamma_\varepsilon \Gamma(1+\varepsilon)$, which is possible for long times but small $\omega$, the expansion $E_\alpha(z) = \sum_{j=0}^{\infty} z^j \Gamma^{-1}(\alpha j + 1)$ must be used in (20), which gives

$$X(t) = \frac{2k_B T}{\gamma_\varepsilon \Gamma(1+\varepsilon)\Gamma(2-\varepsilon)}\left(t^{1-\varepsilon} - \frac{M\omega^2 \Gamma(2-\varepsilon)}{\gamma_\varepsilon \Gamma(1+\varepsilon)\Gamma(3-2\varepsilon)} t^{2-2\varepsilon} + \ldots\right). \quad (21)$$

In what follows we will refer to this case as to the weak trap approximation, while Eq. (19) describes the MSD for a particle in a "strong trap".

By using Eq. (19), we obtain the strong-trap $\langle \phi^2(t) \rangle$ for (9) and (11) that describes the attenuation in the steady-gradient experiment,

$$\langle \phi^2(t) \rangle \approx \frac{2k_B T \gamma_n^2 g^2}{M\omega^2}\left(\frac{t^2}{2} - \frac{\varepsilon}{\varepsilon+1}\frac{\gamma_\varepsilon}{M\omega^2} t^{\varepsilon+1}\right). \quad (22)$$

The weak trap approximation (21) gives

$$\langle \phi^2(t) \rangle = \frac{2k_B T \gamma_n^2 g^2}{\gamma_\varepsilon \Gamma(1+\varepsilon)\Gamma(2-\varepsilon)}\left(\frac{t^{3-\varepsilon}}{3-\varepsilon} - \frac{M\omega^2 \Gamma(2-\varepsilon)}{\gamma_\varepsilon \Gamma(1+\varepsilon)\Gamma(3-2\varepsilon)}\frac{t^{4-2\varepsilon}}{4-2\varepsilon} + \ldots\right). \quad (23)$$

The echo long-time approximation for the strong trap is obtained from Eq. (12),

$$\left\langle \phi^2(\delta, \Delta) \right\rangle \approx -2k_B T \left( \frac{\gamma_n g}{M \omega^2} \right)^2 \varepsilon \gamma_\varepsilon \psi_{\varepsilon-1}, \tag{24}$$

where $\psi_\alpha = [(\Delta+\delta)^{\alpha+2} + (\Delta-\delta)^{\alpha+2} - 2\Delta^{\alpha+2} - 2\delta^{\alpha+2}][(\alpha+1)(\alpha+2)]^{-1}$. At $\varepsilon = 0$ (the standard Langevin case)

$$\left\langle \phi^2(\delta) \right\rangle \approx \frac{4k_B T \gamma \gamma_n^2 g^2}{(M\omega^2)^2} \delta \tag{25}$$

and does not depend on $\Delta$. For the steady-gradient echo at $t = 2\tau$ ($\delta = \Delta = \tau$)

$$\left\langle \phi^2(2\tau) \right\rangle \approx \frac{8(1-2^{\varepsilon-1})\gamma_\varepsilon}{1+\varepsilon} \cdot \frac{k_B T \gamma_n^2 g^2}{(M\omega^2)^2} \tau^{1+\varepsilon} \tag{26}$$

and at $\delta \ll \Delta$

$$\left\langle \phi^2(\delta) \right\rangle \approx \frac{4k_B T \gamma_\varepsilon \gamma_n^2 g^2}{(M\omega^2)^2 (1+\varepsilon)} \delta^{1+\varepsilon}. \tag{27}$$

The influence of the weak trap is given by the second term in $X(t)$, Eq. (21). The full result is

$$\left\langle \phi^2(\delta, \Delta) \right\rangle = \frac{2k_B T \gamma_n^2 g^2}{\gamma_\varepsilon \Gamma(1+\varepsilon) \Gamma(2-\varepsilon)} \left[ \psi_{1-\varepsilon} - \frac{M\omega^2 \Gamma(2-\varepsilon)}{\gamma_\varepsilon \Gamma(1+\varepsilon) \Gamma(3-2\varepsilon)} \psi_{2-2\varepsilon} + \ldots \right], \tag{28}$$

When $\varepsilon = 0$, $\langle \phi^2(\delta, \Delta) \rangle = 2k_B T \gamma_n^2 g^2 \gamma^{-1}[\psi_1 - M\omega^2 \psi_2/2\gamma + \ldots]$ with $\psi_1 = \delta^2(\Delta - \delta/3)$ and $\psi_2 = (\delta\Delta)^2$. The first term, as Eq. (16) at this limit, corresponds to the well-known result [3].

### 6. On interpretation of experiments

The formulas for the NMR signal attenuation based on the solutions of the GLE describing fractional BM considered in the previous sections generalize the results from the standard Langevin theory for both free particles and particles trapped in a harmonic potential. At long times, which are of main interest from the point of view of experiments, the free particles within this model undergo subdiffusion, and the MSD of the trapped particles is, see Eq. (20), $X(t) \approx 2k_B T / M\omega^2$. In the latter case, as it is seen from Eq. (22), the NMR signal in the steady gradient experiment can be significantly influenced by the particle motion, while such $X(t)$ (time-independent) does not affect the signal of the pulsed-gradient echo at all.

In what follows we describe the experiments in which subdiffusion of water in human brain tissues [23] has been clearly demonstrated. The fractional BM thus seems to be one of the possible models to be used in interpretation of these experiments. The choice of this system is important also because of the fact that the diffusion in neuronal tissues has been associated with alterations in physiological and pathological states and its understanding can have important implications.

By using Eq. (43) [23] (see above the text after Eq. (16)), in [23] Eq. (44) was at $\Delta = \delta = t$ used in the form

$$\ln[S(t)/S(0)] = 2[(3-\varepsilon)\Gamma(2-\varepsilon)]^{-1} \gamma_n^2 g^2 D t^{3-\varepsilon}. \qquad (29)$$

The correct expression should however read, from Eq. (16),

$$-\ln[S(t)/S(0)] \approx 4(2^{1-\varepsilon}-1)[(2-\varepsilon)(3-\varepsilon)\Gamma(2-\varepsilon)\gamma_\varepsilon]^{-1} \gamma_n^2 g^2 k_B T t^{3-\varepsilon}. \qquad (30)$$

Along with a different non-dimensional coefficient, (29) contains the parameter $D$, which is not the diffusion coefficient of the dimension $L^2 T^{-1}$ as in [23], but should be $D_\varepsilon = k_B T / \gamma_\varepsilon$, which dimension is $L^2 T^{-1+\varepsilon}$. The determination of $D$ and $\alpha = 1 - \varepsilon$ in [23] from experiments is thus flawed. So, even if $\alpha$ was correctly extracted from the experiments, the numerical value of $D$ found from the measured $\ln[S(t)/S(0)]$ would have to be divided by $2(2^\alpha - 1)/(\alpha+1)$, which for $\alpha \in [0, 1]$ changes from 0 to 1 (e.g., if $\alpha = 0.5$, see Table II [23], $D_\varepsilon$ should be about 1.8 times larger than $D$). The parameters $\alpha$ and $D_\varepsilon$ of the fractional BM model are easily obtainable from the comparison of the presented theory with experimental data. Such a comparison is illustrated by Fig. 1 that shows the time dependence of $S(t)$ calculated from Eqs. (30) and (29) [57] for the same $\varepsilon$ as in [23].

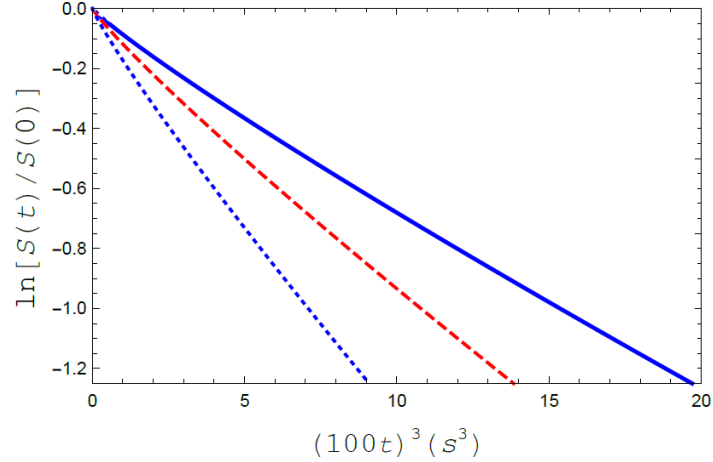

**Fig. 1.** Attenuation function for the steady-gradient spin-echo signal from water molecules anomalously diffusing in a human brain tissue. The parameters correspond to the experiment ([23], Table I, Fig. 1) and $\varepsilon = 0.31$. The dashed line (red online) is calculated from Eq. (29) with $(\gamma_n g)^2 D_\varepsilon = 3.48 \times 10^4$ s$^{-3+\varepsilon}$ and very well corresponds to the experiment [23] and to Eq. (30), but if $D_\varepsilon$ is $(2-\varepsilon)/[2(2^{1-\varepsilon}-1)] \doteq 1.38$ times larger. The blue lines illustrate the time dependence of the attenuation from Eq. (30) for $(\gamma_n g)^2 D_\varepsilon = 3.48 \times 10^4$ s$^{-3+\varepsilon}$ (full line) and $(\gamma_n g)^2 D_\varepsilon = 2 \times 3.48 \times 10^4$ s$^{-3+\varepsilon}$ (dotted line).

The use of both (29) and (30) lead to a good agreement with the experiment [23], but, for a given $\varepsilon$, at very different values of $D_\varepsilon$. If we assume that the numerical value of $D_\varepsilon$ was correctly determined in [23], an agreement with the experiment is reached for a very distinct $\varepsilon$. For example, if $(\gamma_n g)^2 D_\varepsilon = 3.48 \times 10^4$ s$^{-3+\varepsilon}$, $\varepsilon$ would be approximately 0.42 instead of 0.31 found in [23]. Note that the fit to the experimental data performed in [23] was very accurate (the chi-squared values for goodness of the fit were $< 10^{-5}$). On the other hand, our calculations represented by the dashed line in Fig. 1 very well correspond to this fit: when the difference between the fit (and thus also the data [23]) and our calculation is maximal, the relation of $S(t)$ to that from [23] is 1.01. For all the experimental times our calculations thus show a very good agreement with the data.

We are not aware about a suitable experiment on trapped particles in a fractal environment. For such particles we thus give only an illustration how the particle motion at long times (when $\gamma_\varepsilon \Gamma(\varepsilon+1) t^{\varepsilon+1} / M \gg 1$) influences the steady-gradient echo signal in the cases of strong and weak traps, which requires the conditions $M\omega^2 t^{1-\varepsilon} \gg \gamma_\varepsilon \Gamma(1+\varepsilon)$ and $M\omega^2 t^{1-\varepsilon} \ll \gamma_\varepsilon \Gamma(1+\varepsilon)$ to be fulfilled, respectively. A problem arises that for such illustration we need to know the generalized friction coefficient $\gamma_\varepsilon$ with $\varepsilon \neq 0$. This can be solved by introducing the following relation between $\gamma_\varepsilon = k_B T / D_\varepsilon$ and the usual friction coefficient $\gamma = k_B T / D$ when $\varepsilon = 0$:

$$\gamma_\varepsilon = (k_B T)^\varepsilon R^{-2\varepsilon} \gamma^{1-\varepsilon} . \tag{31}$$

This relation or the equivalent one, $D_\varepsilon = D^{1-\varepsilon} R^{2\varepsilon}$, can be obtained by dimensional consideration. Of course, it is not the most general one. One could assume that $D_\varepsilon = a_\varepsilon + b_\varepsilon D^{1-\varepsilon} R^{2\varepsilon}$, where the term $a_\varepsilon$ has the dimension L$^2$ T$^{-1+\varepsilon}$ and $b_\varepsilon$ is dimensionless. Both these quantities characterize the fractal environment for the Brownian particle and do not depend on its radius $R$. Let us assume that $R$ decreases. The second term in $D_\varepsilon$, which is proportional to $R^{3\varepsilon-1}$ is expected to grow. This is possible if $\varepsilon < 1/3$, i.e., not $\varepsilon < 1$, as initially assumed in the fractional BM model. Moreover, with the increase of $R$ the fractal property of the environment fades away (the fractality is manifested weaker) and $D_\varepsilon$ gets closer to $D$. Thus, the term $a_\varepsilon$ can be assumed negligible and $b_\varepsilon$ equal to 1 or close to it, so that the use of (31) seems to be reasonable. Figures 2 and 3 illustrate the dependence of the attenuation function on $\varepsilon$ for the steady-gradient spin-echo signals for strong and weak traps, calculated from Eq. (24) and (28), respectively.

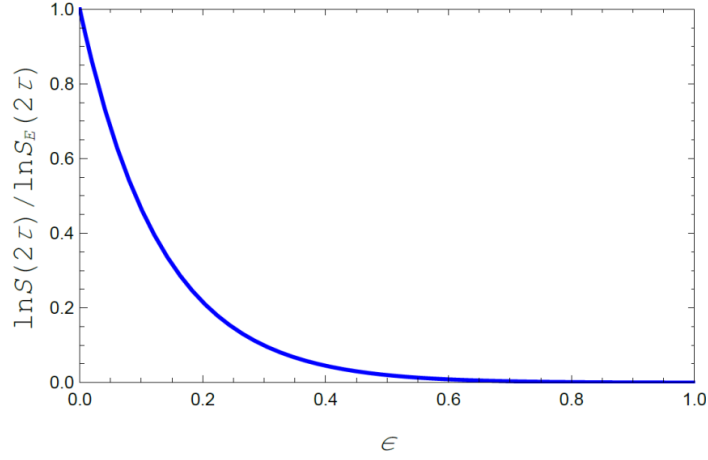

**Fig. 2.** Normalized attenuation function of the steady-gradient echo signal from particles in a strong trap at long times, calculated from Eq. (24) for the model (31), as described in the text. The used parameters are $k = 10^2$ µN/m [58, 59], $\tau = 10$ ms, the viscosity is as for water at $T = 300$ K, when for particles with $R = 1$ µm the friction and diffusion coefficients are $\gamma \approx 16 \cdot 10^{-9}$ kg/s and $D \approx 0.26 \cdot 10^{-12}$ m$^2$/s, and the mass $M \approx 4.2 \cdot 10^{-15}$ kg, assuming the particles' density close to that of water.

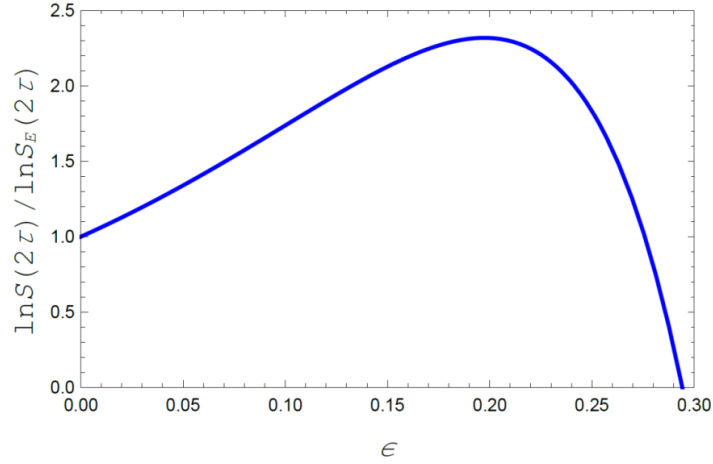

**Fig. 3.** Normalized attenuation function of the steady-gradient echo signal from particles in a weak trap at long times, calculated from Eq. (28) for the model (31), as described in the text. The trap stiffness is small, $k = 1$ µN/m [60], the viscosity is as for blood at $T = 300$ K, $R = 1$ µm, $\gamma = 64 \cdot 10^{-9}$ kg/s, $D = 6.47 \cdot 10^{-14}$ m$^2$/s, and $\tau = 10^{-2}$ s.

The results for $\ln S(2\tau)$ are normalized to $\ln S_E(2\tau)$, where $S_E(2\tau) = \exp(-2k_B T \gamma_n^2 g^2 k^{-2} \gamma \tau)$ follows coming from the Einstein theory of diffusion (see Eqs. (25) and (26) at $\varepsilon = 0$). The increase of the fractal parameter $\varepsilon$ leads to a significant difference from the attenuation in the case of normal diffusion. For the strong trap the echo signal disappears when $\varepsilon$ becomes close to 1. In Fig. 3 only small values of $\varepsilon$ are relevant, since with the growth of $\varepsilon$ the weak-trap approximation in (28) becomes inappropriate.

## 7. Conclusion

In conclusion, we have shown how the NMR experiments on particles exhibiting the Brownian motion with memory can be interpreted. This allows, in particular, obtaining characteristics of the particles' environment. In this paper the dynamics of the particles is assumed to be governed by the generalized Langevin equation for the fractional Brownian motion. The particles can be free or trapped in a harmonic potential, which can be real or used to simulate the motion in pores (restricted Brownian motion) [61]. Although various NMR methods serve as accurate probes for characterization of the random motion of spin-bearing particles in different systems, the interpretation of the NMR experiments in the literature is limited to long times (when the particles are in the diffusion regime) or to the stochastic motion without memory. Nowadays it is well known that in the description of the behavior of many systems these simplifications fail. In this work we apply new formulas for the attenuation function of the NMR signals that can be directly used to describe the behavior of particles undergoing any kind of the stochastic dynamics not only in the diffusion limit, but at any times. The formulas just require the knowledge of the mean square displacement of the particle. For the considered model it is obtained by an exceedingly simple way, which is appropriate for other generalized Langevin models as well. Although our main aim was to give a description of the dynamics of Brownian particles much larger than the surrounding molecules, the presented approach leads to a good agreement with the experiments on water molecules freely diffusing in neuronal tissues [23]. However, the fractal parameter $\varepsilon$ and the generalized diffusion coefficient $D_\varepsilon$ are shown to be very different from those determined in the original experiments. Finally, the influence of weak and strong harmonic traps on the attenuation function was considered in detail. This required modeling the generalized friction coefficient of the Brownian particles in a fractal environment. Based on the dimensional consideration and expected properties of $D_\varepsilon$, we suggest that $\varepsilon$ should be smaller than 1/3, instead of $\varepsilon < 1$, as it is commonly assumed in the fractional Brownian motion model. Except for the description of diffusion of molecules in biological tissues [23] (see also [18, 32]), the presented theory could be applied in interpretations of the NMR experiments on transport in heterogeneous systems such as the network of pores filled by a solvent [13–15], NMR measurements of the anomalous Brownian motion in microemulsions and polymers [3, 32, 62] , chemical reaction kinetics [16, 63], distance fluctuations of particles in complex fluids, e.g., liquid crystals showing in the vicinity of the isotropic-to-nematic transition a power-law decay profiles in the relaxation behavior [64], dynamics of random-coil macromolecules in sufficiently concentrated solution (gels) [19, 65], anomalous Brownian motion phenomenon in crowded environments, such as living cells [24], and protein conformational dynamics [66, 67].

## Acknowledgment

This work was supported by the Ministry of Education and Science of the Slovak Republic through grant VEGA No. 1/ 0348/15.